\documentclass[prl, aps, floatfix, twocolumn, superscriptaddress, preprintnumbers]{revtex4-1}

\usepackage[colorlinks = true, citecolor = blue, linkcolor = blue, urlcolor = blue]{hyperref}

\usepackage{graphicx}

\usepackage{mathtools}
\usepackage{amsmath}
\usepackage{amssymb}
\usepackage{bm}
\usepackage{mathrsfs}
\usepackage{slashed}
\usepackage{physics}
\usepackage{xcolor}
\usepackage{diagbox}
\allowdisplaybreaks
\usepackage[capitalise]{cleveref}

\usepackage{atbegshi,picture}

\begin{document}
\title{Polarized jet anisotropy at the Electron-Ion Collider}

\author{Zhong-Bo Kang}
\email{zkang@physics.ucla.edu}
\affiliation{Department of Physics and Astronomy, University of California, Los Angeles, CA 90095, USA}
\affiliation{Mani L. Bhaumik Institute for Theoretical Physics, University of California, Los Angeles, CA 90095, USA}
\affiliation{Center for Frontiers in Nuclear Science, Stony Brook University, Stony Brook, NY 11794, USA}

\author{Hongxi Xing}
\email{hxing@m.scnu.edu.cn}
\affiliation{State Key Laboratory of Nuclear Physics and Technology, Institute of Quantum Matter, South China Normal University, Guangzhou 510006, China}
\affiliation{Guangdong Basic Research Center of Excellence for Structure and Fundamental Interactions of Matter, Guangdong Provincial Key Laboratory of Nuclear Science, Guangzhou 510006, China}
\affiliation{Southern Center for Nuclear-Science Theory (SCNT), Institute of Modern Physics, Chinese Academy of Sciences, Huizhou 516000, China}

\author{Fanyi Zhao}
\affiliation{Center for Theoretical Physics – a Leinweber Institute, Massachusetts Institute of Technology, Cambridge, MA 02139, USA}

\author{Yiyu Zhou}
\email{zyy@impcas.ac.cn}
\affiliation{Institute of Modern Physics, Chinese Academy of Sciences, Lanzhou, Gansu 730000, China}
\affiliation{University of Chinese Academy of Sciences, Beijing 100049, China}

\begin{abstract}
Jets provide a powerful probe of the three-dimensional spin structure of the nucleon, a central goal of the Electron-Ion Collider. Yet the observed jet defines an axis that breaks the azimuthal isotropy of soft-gluon radiation, thereby reshaping the very asymmetries used to extract that structure. Using transverse-momentum-dependent (TMD) QCD factorization, we show for the first time that this jet-induced anisotropy imposes a parity selection rule on polarized asymmetries. Expanding the transversely polarized structure functions in harmonics $\cos(n\phi_{qJ})$, where $\phi_{qJ}$ is the angle between the jet and the lepton-jet momentum imbalance, makes this rule explicit: the symmetry of each harmonic is fixed by the parity of $n$, independently of the magnitudes of the harmonic coefficients. For the Sivers function, the canonical left-right asymmetry about the proton spin survives for even $n$ but is replaced by a counterintuitive left-right symmetry for odd $n$. The worm-gear function retains its up-down asymmetry at every harmonic while being left-right symmetric for even $n$ and asymmetric for odd $n$. At EIC kinematics, the higher harmonics studied here are predicted to be individually measurable, providing new observables for the azimuthal dynamics of soft radiation and an essential ingredient in precision extractions of nucleon structure.
\end{abstract}

\maketitle

\textit{Introduction.---}Three-dimensional (3D) imaging of the nucleon is a central frontier in modern hadron physics \cite{Boussarie:2023izj}, forming a primary scientific pillar of the future Electron-Ion Colliders (EIC)~\cite{AbdulKhalek:2021gbh, Anderle:2021wcy}. This 3D structure is fundamentally encoded in transverse momentum-dependent parton distribution functions (TMD PDFs) \cite{Collins:2003fm, Barone:2010zz, Aybat:2011zv, Constantinou:2020hdm}. Historically, these distributions have been extracted using semi-inclusive deep inelastic scattering (SIDIS) \cite{Mulders:1995dh, Anselmino:2005ea, Anselmino:2005nn, Vogelsang:2005cs, Collins:2005ie, Collins:2005wb, Anselmino:2006yc, Bacchetta:2006tn, Anselmino:2007fs, Anselmino:2008sga, Barone:2009hw, Kang:2012xf, Anselmino:2012aa, Anselmino:2013vqa, Signori:2013mda, Anselmino:2013lza, Echevarria:2014xaa, Lefky:2014eia, Kang:2015msa, Anselmino:2015sxa, Anselmino:2016uie, Bacchetta:2017gcc, Lin:2017stx, Scimemi:2019cmh, DAlesio:2020vtw, Cammarota:2020qcw, Echevarria:2020hpy, Bury:2020vhj, Bury:2021sue, Bhattacharya:2021twu, Gamberg:2022kdb, Bacchetta:2022awv, Horstmann:2022xkk, Alrashed:2023xsv, Boglione:2024dal, Yang:2024bfz, Bacchetta:2024qre, Moos:2025sal}, where polarized TMDs imprint nontrivial angular correlations that link the parton's transverse motion to the nucleon's spin \cite{DAlesio:2007bjf, Boer:2011fh, Aidala:2012mv, Boussarie:2023izj}. 

More recently, jet production has emerged as a high-precision, complementary probe, often termed ``jet tomography'' \cite{Kogler:2018hem,Larkoski:2017jix,AbdulKhalek:2022hcn, Abir:2023fpo}. For instance, back-to-back dijet and lepton-jet production in polarized collisions offer direct access to spin-dependent distributions like the Sivers function \cite{Boer:2003tx, Bacchetta:2005rm, Vogelsang:2005cs, Bomhof:2007su, Qiu:2007ar, Qiu:2007ey, Vogelsang:2007jk, Boer:2009nc, Liu:2020jjv, Kang:2020xez}.
However, recent theoretical advancements have revealed that jet observables harbor subtle QCD dynamics that complicate this standard picture. Unlike inclusive single-hadron production, the detection of a final-state jet defines a specific axis in the transverse plane. This axis explicitly breaks the spatial uniformity of soft-gluon radiation \cite{Buffing:2018ggv, Hatta:2020bgy}. In unpolarized scattering, this symmetry breaking induces a measurable azimuthal anisotropy, typically yielding a ${\rm cos}(2\phi_{qJ})$ harmonic modulation, driven entirely by soft-gluon radiation effects \cite{Dumitru:2015gaa, Dominguez:2010xd, Metz:2011wb, Dominguez:2011br, Dumitru:2016jku, Boer:2016fqd, Dumitru:2018kuw, Zhao:2021kae, Boussarie:2021ybe, Caucal:2022ulg, Hatta:2021jcd, Tong:2022zwp, Tong:2023bus}. While this phenomenon complicates the extraction of unpolarized distributions, its implications for \textit{polarized} scattering, where azimuthal asymmetries are the very signals of nucleon spin structure, have remained critically unexplored.

In this Letter, we demonstrate for the first time that jet-induced soft-gluon radiation anisotropy fundamentally reshapes the interpretation of polarized nucleon structure. Within TMD QCD factorization for back-to-back lepton-jet production in polarized $e+p$ collisions, we show that the azimuthal dependence of the soft function generates an array of nontrivial higher-harmonic patterns. Conventionally, the Sivers effect is treated strictly as a left-right asymmetry relative to the nucleon spin axis \cite{STAR:2008ixi, She:2008tu, Sun:2009ew, Gamberg:2014eia, GrossePerdekamp:2015xdx, Huang:2015vpy}. We find that this picture is preserved by the even harmonics, whereas the odd harmonics generate a counterintuitive left-right \textit{symmetry}. Similarly, the worm-gear function, long recognized as the source of up-down spin asymmetry \cite{Huang:2015vpy}, retains this asymmetry at every harmonic while being left-right symmetric for even harmonics and asymmetric for odd ones.

These jet-induced modulations produce distinct angular patterns. We demonstrate that these higher harmonics produce sizable, measurable deviations from canonical expectations at EIC kinematics. Consequently, interpretative frameworks that neglect soft-function azimuthal dependence risk mistaking these jet-induced QCD dynamics for fundamental nucleon structure. Resolving these harmonic signatures is therefore essential for precision TMD tomography at the future EIC.

\textit{QCD factorization formalism.---}%%
We consider lepton-jet production in inclusive lepton-proton scatterings:
\begin{align}
e \pqty{\ell, \lambda_e} + p \pqty{P, S} \to e \pqty{\ell'} + \mathrm{jet} \pqty{p_J} + X,
\end{align}
where $\ell$ and $P$ denote the momenta of the incoming lepton and proton, with their corresponding helicities (or spins) given by $\lambda_e$ and $S$, respectively. The scattered-lepton and jet momenta are denoted by $\ell'$ and $p_J$, with transverse components $\boldsymbol{\ell}'_T$ and $\boldsymbol{p}_T$, respectively.

\begin{figure}[hbt]
\centering
\includegraphics[width = 0.95 \columnwidth]{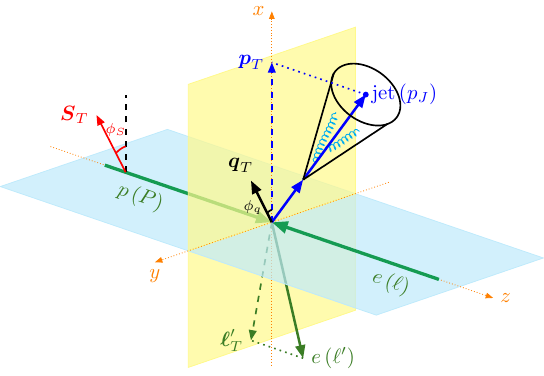}
\caption{Kinematic configuration for back-to-back lepton-jet production in $ep$ collisions, where the jet azimuthal angle is defined to be aligned with the $x$-axis.
}
\label{f.kinematics}
\end{figure}

Working within the one-photon-exchange approximation and neglecting lepton masses, the cross section can be written in terms of the structure functions \cite{Kang:2021ffh}:
\begin{align} \label{e.full_structure_function}
\frac{\dd{\sigma}}{\dd{\mathcal{PS}}}
& =
F_{UU} + \lambda_e \lambda_p F_{LL} + S_T
\Big[
\sin(\phi_q - \phi_S) F^{\sin(\phi_q - \phi_S)}_{UT}
\nonumber \\
& \quad +
\lambda_e \cos(\phi_q - \phi_S) F^{\cos(\phi_q - \phi_S)}_{LT}
\Big],
\end{align}
where the indices in the structure functions $F_{AB}$ ($A,B\in\{U,L,T\}$) specify the polarization states of the incoming lepton and proton.
The phase space is given as $\dd{\mathcal{PS}} \equiv \dd{p_T^2} \dd{y_J} q_T \dd{q_T} \dd{\phi_q}$, $\boldsymbol{q}_T \equiv\boldsymbol{\ell}'_T + \boldsymbol{p}_T$ is the transverse momentum imbalance, and $\phi_q$ is its azimuthal angle.
We orient the $x$-axis along the jet direction such that $\phi_J = 0$, and $\phi_S$ denotes the azimuthal angle of $\boldsymbol{S}_T$.
The kinematics are illustrated in \cref{f.kinematics}. 

In this Letter, we focus on the back-to-back lepton-jet configuration in the azimuthal plane, a regime highly sensitive to proton TMDs. In the correlation limit $\abs{\boldsymbol{q}_T} \ll \abs{\boldsymbol{p}_T}$, it has been shown that the unpolarized structure functions $F_{UU}$ can be factorized into a convolution of a hard function, a TMD PDF, a jet function, a global soft function, and a collinear-soft function \cite{Kang:2021ffh}. Following a similar strategy, we derive the longitudinally polarized structure function $F_{LL}$ by replacing the unpolarized TMDs with helicity-dependent TMDs, offering a sensitive probe of nucleon helicity structure.
 
Within TMD factorization, for unpolarized leptons scattering off a transversely polarized proton,
$F_{UT}^{\sin(\phi_q - \phi_S)}$ factorizes in coordinate $\boldsymbol{b}$-space as:
\begin{align}
\label{e.Sivers_cs}
F_{UT}^{\sin(\phi_q - \phi_S)}
& =
\hat{\sigma}_0 H \pqty{Q, \mu} M 
\sum_q e_q^2 \mathcal{J}_q \pqty{p_T R, \mu}
\int \frac{b \dd{b}}{2 \pi}
\nonumber \\
& \times
x \widetilde{f}_{1T, q/p}^{\perp, \pqty{1}} \pqty{x, b^2, \mu, \zeta}
\int \frac{\dd{\phi_{bq}}}{2 \pi}
e^{i \boldsymbol{q}_T \cdot \boldsymbol{b}}
\nonumber \\
& \times
i b \cos(\phi_{bq})
S_{\mathrm{global}} \pqty{\boldsymbol{b}, \mu}
S_{\mathrm{cs}} \pqty{\boldsymbol{b}, R, \mu}
,
\end{align}
where $M$ is the proton mass, $R$ is the jet radius, and $\phi_{bq}$ is the relative azimuthal angle between the vectors $\boldsymbol{b}$ and $\boldsymbol{q}_T$.
$\hat{\sigma}_0$ is the Born-level partonic cross section for unpolarized $eq \to eq$ scattering \cite{Kang:2021ffh}.
$H \pqty{Q, \mu}$ is the hard function describing perturbative corrections to the partonic scattering, and $\mathcal{J}_q \pqty{p_T R, \mu}$ is the quark jet function characterizing the production of the outgoing jet \cite{Ellis:2010rwa, Liu:2018trl, Arratia:2020nxw}. The collinear-soft function $S_{\mathrm{cs}} \pqty{\boldsymbol{b}, R, \mu}$ (cyan helices in \cref{f.kinematics}) governs soft radiation that resolves the jet cone and depends on the jet radius $R$. The global soft function $S_{\mathrm{global}} \pqty{\boldsymbol{b}, \mu}$ describes wide-angle soft radiation without phase-space restrictions and does not resolve the jet cone. Both soft functions depend on the full transverse vector $\boldsymbol{b}$, rather than only on its magnitude. The term $\widetilde{f}_{1T, q/p}^{\perp, \pqty{1}} \pqty{x, b^2, \mu, \zeta}$ is the quark Sivers function in $\boldsymbol{b}$-space, $\mu$ is the renormalization scale, and $\zeta$ is the Collins-Soper parameter \cite{Collins:2011zzd, Ebert:2019okf}. The one-loop expressions for these functions are given in Refs.~\cite{Buffing:2018ggv, Kang:2021ffh, delCastillo:2020omr}, and we have verified the RG consistency of the factorized expression and the cancellation of regulator dependence.

Crucially, this vector dependence makes both soft functions in \cref{e.Sivers_cs} sensitive to $\phi_{bJ}$, the azimuthal angle of $\boldsymbol{b}$ relative to the observed jet axis. At one loop, this angular dependence enters through logarithms involving $-2i\cos{\phi_{bJ}}$.
Here, we investigate for the first time how this azimuthal angle dependent soft gluon radiation contributes to polarized scattering, altering spin asymmetries and the interpretation of 3D nucleon imaging using jet tomography. In particular, we use the derivative of the Jacobi-Anger identity (the arguments $q_T b$ in the Bessel functions are omitted for brevity) to expand the azimuthal dependence:
\begin{align}
& \quad i b \cos(\phi_{bq}) e^{i b q_T \cos(\phi_{bq})}
\nonumber \\
& =
b
\bqty{
-J_1
+
\sum_{n=1}^{\infty} i^n \pqty{J_{n-1} - J_{n+1}} \cos(n \phi_{bq})
} .
\end{align}
This allows us to expand $F_{UT}^{\sin(\phi_q - \phi_S)}$ as a harmonic series in $\cos(n \phi_{qJ})$:
\begin{align}
\label{e.F_UT_expanded}
F_{UT}^{\sin(\phi_q - \phi_S)}
=
\sum_{n=0}^{\infty} A_n^{\mathrm{Sivers}} \cos(n \phi_{qJ})
,
\end{align}
where $\phi_{qJ}$ is the azimuthal angle difference between $\boldsymbol{q}_T$ and $\boldsymbol{p}_T$. The harmonic coefficients $A_n^{\mathrm{Sivers}}$ for $n \geqslant 0$ factorize as:
\begin{align}
A_n^{\mathrm{Sivers}}
& \equiv
\hat{\sigma}_0 H \pqty{Q, \mu}  M
\sum_q e_q^2 \mathcal{J}_q \pqty{p_T R, \mu}
\int \frac{b^2 \dd{b}}{2 \pi} x
\label{e.A_n_Sivers} \nonumber\\
&\times
\widetilde{f}_{1T, q/p}^{\perp, \pqty{1}} \pqty{x, b^2, \mu, \zeta}
i^n \pqty{\frac{1}{2}}^{\delta_{0n}}\Big(J_{n-1} - J_{n+1}\Big)
\nonumber \\
&\times
\int \frac{\dd{\phi_{bJ}}}{2 \pi}
\cos(n \phi_{bJ})
S_{\mathrm{global}} \pqty{\boldsymbol{b}, \mu}
S_{\mathrm{cs}} \pqty{\boldsymbol{b}, R, \mu}
, 
\end{align}
where $\delta_{0n}$ is the Kronecker delta. While harmonic flow patterns shown in \cref{e.F_UT_expanded} have been identified in unpolarized collisions \cite{Hatta:2021jcd, Tong:2022zwp}, we present the first theoretical demonstration of these anisotropies in polarized scattering and quantify their impact on spin asymmetries and the nucleon 3D structure.

Similarly, for transversely polarized proton scattering with a longitudinally polarized lepton, the cross section depends on the worm-gear function $g_{1T}$ embedded within the structure function $F_{LT}^{\cos(\phi_q - \phi_S)}$. Because soft radiation is spin-independent, the factorization for $F_{LT}^{\cos(\phi_q - \phi_S)}$ can be derived directly from $F_{UT}^{\sin(\phi_q - \phi_S)}$ in \cref{e.Sivers_cs} via the substitutions $\hat{\sigma}_0 \to \hat{\sigma}_L$ and $\widetilde{f}_{1T, q/p}^{\perp, \pqty{1}} \to \widetilde{g}_{1T, q/p}^{\pqty{1}}$. Here, $\widetilde{g}_{1T, q/p}^{\pqty{1}}$ is the coordinate-space worm-gear function \cite{Tangerman:1994eh, Kotzinian:1995cz}, and $\hat{\sigma}_L$ represents the Born-level partonic cross section for $e_L q_L \to eq$. The hard, global soft, and collinear-soft functions remain identical to the Sivers case. Consequently, we can decompose $F_{LT}^{\cos(\phi_q - \phi_S)}$ as:
\begin{align}
F_{LT}^{\cos(\phi_q - \phi_S)}
=
\sum_{n=0}^{\infty} A_n^{\textrm{worm-gear}} \cos(n \phi_{qJ})
, \label{e.F_LT_expanded}
\end{align}
where the coefficients $A_n^{\textrm{worm-gear}}$ are defined identically to \cref{e.A_n_Sivers}, upon applying the aforementioned substitutions.

The origin of these symmetry patterns can be seen directly at the
cross-section level. For the orientation used below, $\phi_J = 0$ and
$\phi_S = \pi/2$, so that $\phi_{qJ} = \phi_q$, the angular weights
multiplying the $n$th Sivers and worm-gear harmonics are:
\begin{align}
\mathcal{W}_{UT}^{(n)}
& \equiv \sin(\phi_q - \phi_S) \cos(n \phi_{qJ})
\nonumber \\
& = - \frac{1}{2}
\bqty{
\cos\big( (n+1) \phi_q \big) + \cos\big( (n-1) \phi_q \big)
} ,
\label{e.Sivers_selection_rule}
\\
\mathcal{W}_{LT}^{(n)}
& \equiv \cos(\phi_q - \phi_S) \cos(n \phi_{qJ})
\nonumber \\
& = \frac{1}{2}
\bqty{
\sin\big( (n+1) \phi_q \big) - \sin\big( (n-1) \phi_q \big)
} .
\label{e.wormgear_selection_rule}
\end{align}
Under left-right reflection about the spin axis, $\phi_q \to \pi - \phi_q$, these weights obey $\mathcal{W}_{UT}^{(n)} \to (-1)^{n+1} \mathcal{W}_{UT}^{(n)}$ and $\mathcal{W}_{LT}^{(n)} \to (-1)^{n} \mathcal{W}_{LT}^{(n)}$, so the Sivers harmonics are left-right asymmetric for even $n$ and symmetric for odd $n$, while the worm-gear harmonics show the opposite pattern. Under up-down reflection, $\phi_q \to - \phi_q$, one has $\mathcal{W}_{LT}^{(n)} \to - \mathcal{W}_{LT}^{(n)}$ for every $n$, so the canonical worm-gear up-down asymmetry survives at all harmonics. In addition, the odd-$n$ worm-gear harmonics generate a left-right asymmetry with no counterpart in the standard SIDIS worm-gear asymmetry. These parity selection rules hold for all $n$, independently of the magnitudes of the coefficients $A_n$.

\textit{Phenomenology at EIC.---}%%
We now present numerical predictions for the anisotropic observables associated with both the Sivers and worm-gear functions. To quantify each harmonic series, we define the associated jet anisotropy as the $\cos[2](n \phi_{qJ})$-weighted average of the asymmetry, namely:
\begin{align}
R_{UT}^{\expval{\cos(n \phi)}}
\equiv
\int_0^{2 \pi} \frac{\dd{\phi_{qJ}}}{2\pi} \cos[2](n \phi_{qJ}) \frac{A_n^{\mathrm{Sivers}} }{A_0^{\mathrm{Sivers}}}
,
\end{align}
with $R_{LT}^{\expval{\cos(n \phi)}}$ defined analogously. Here, $\phi$ in the superscript is shorthand for $\phi_{qJ}$. Normalizing each channel to its $n=0$ coefficient quantifies the size of the higher harmonics relative to the conventional Sivers or worm-gear contribution.

\begin{figure}[tb]
\centering
\includegraphics[width = 0.9 \columnwidth]{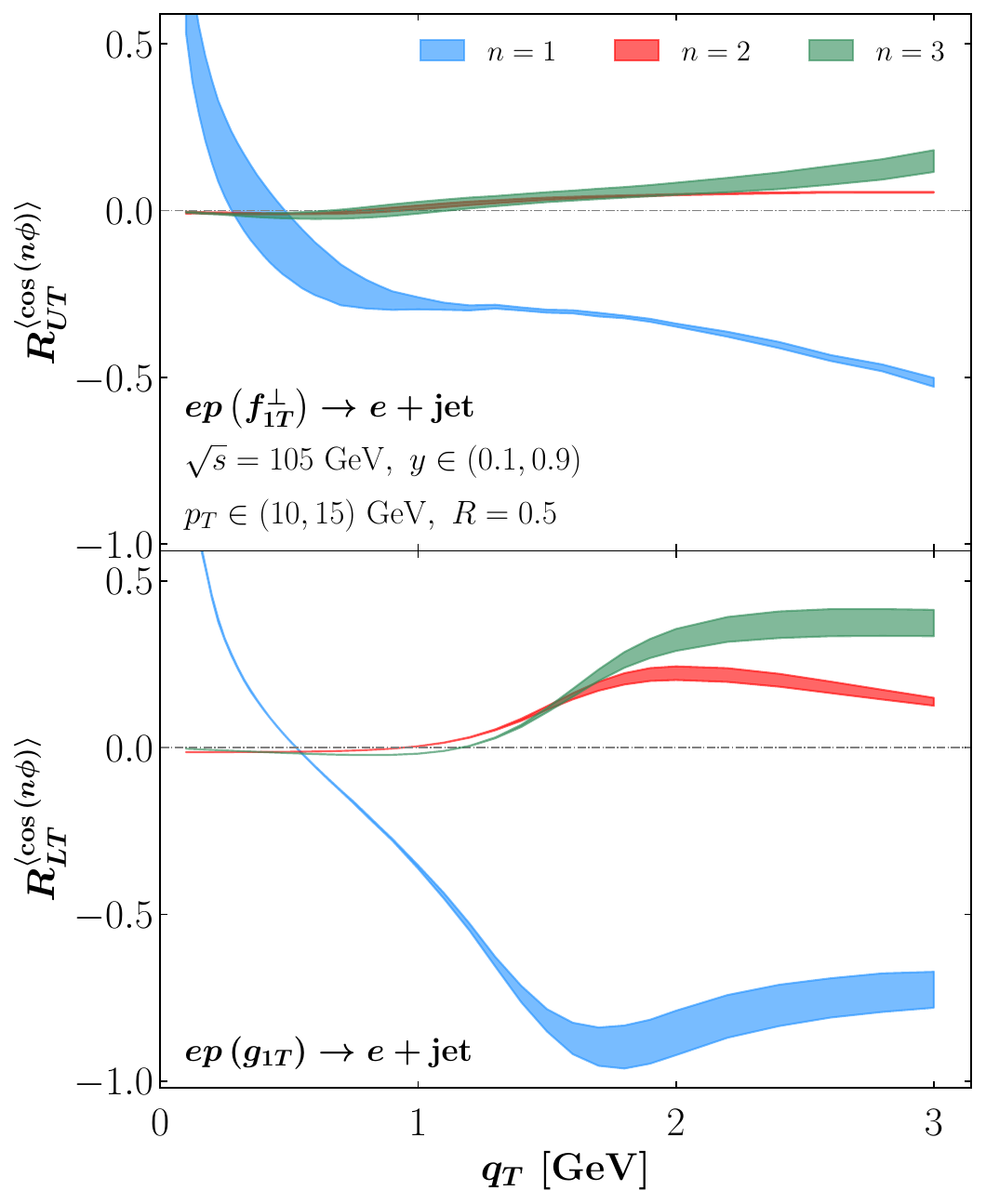}
\caption{\
The polarized azimuthal anisotropy $R_{UT}^{\expval{\cos(n \phi)}}$ and $R_{LT}^{\expval{\cos(n \phi)}}$ related to the Sivers function $f_{1T}^{\perp}$ \cite{Echevarria:2020hpy} and worm-gear function $g_{1T}$ \cite{Bhattacharya:2021twu}, with EIC kinematics at $\sqrt{s} = 105~\mathrm{GeV}$, event inelasticity $y \in \pqty{0.1, 0.9}$, $10 < p_T < 15~\mathrm{GeV}$, $0 < q_T < 3~\mathrm{GeV}$ and jet radius $R = 0.5$.
}
\label{f.radial_anisotropy}
\end{figure}

In \cref{f.radial_anisotropy}, we plot the Sivers jet anisotropy observables (upper panel) and the worm-gear jet anisotropy (lower panel) for a jet radius of $R = 0.5$. The observables $R_{UT}^{\expval{\cos(n \phi)}}$ and $R_{LT}^{\expval{\cos(n \phi)}}$ exhibit qualitatively similar behavior but differ in magnitude. In both scenarios, the $n=1$ harmonic is dominant across the considered $q_T$ range, except near a zero-crossing at $q_T \sim 0.5~\mathrm{GeV}$, and becomes negative in both cases in large $q_T$ region. Importantly, the magnitudes for the $n = 1, 2,$ and $3$ harmonics are large enough to be measured at EIC kinematics, making them compelling targets for future experiments. In particular, $R_{LT}^{\expval{\cos(\phi)}}$ ($n=1$) is strikingly sizable, especially near $q_T \sim 2~\mathrm{GeV}$. This enhancement originates from the Bessel function modulation inside the convolution integral in \cref{e.A_n_Sivers}.

\begin{figure}
\centering
\includegraphics[width = 0.98 \columnwidth]{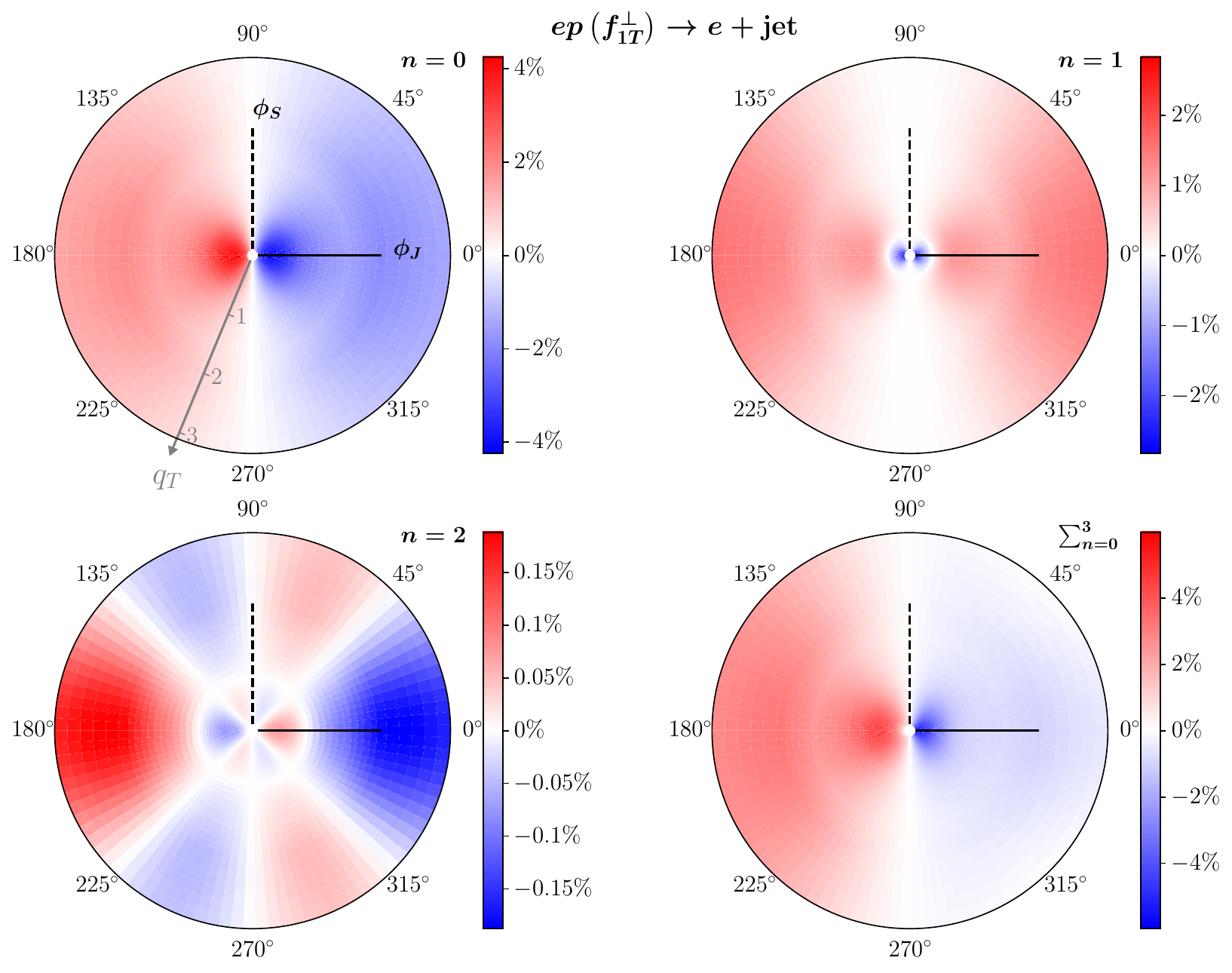}
\caption{
Heat map for the components of polarized azimuthal anisotropy $S_T \sin(\phi_q - \phi_S) \cos(n \phi_{qJ}) A_n^{\mathrm{Sivers}} / A_0^{\mathrm{unp.}}$ related to the Sivers function $f_{1T}^{\perp}$ with the same kinematics as in \cref{f.radial_anisotropy}.
The radial direction is $q_T$ and the angular direction is $\phi_q$ of $\boldsymbol{q}_T$.
We have set $\phi_J$ (black solid line) at 0 and $\phi_S$ of the proton transverse spin $\boldsymbol{S}_T$ at $\pi / 2$ (black dashed line).
}
\label{f.Sivers_heat_map}
\end{figure}

To visualize the impact of these higher harmonics, we map the normalized Sivers contributions $S_T \sin(\phi_q - \phi_S) \cos(n \phi_{qJ}) A_n^{\mathrm{Sivers}} / A_0^{\mathrm{unp.}}$ from \cref{e.F_UT_expanded} as functions of both $q_T$ and the azimuthal angle $\phi_q$ in \cref{f.Sivers_heat_map}, fixing $\phi_J = 0$ and $\phi_S = \pi / 2$. Unlike the ratios defined above, the heat maps normalize the polarized contributions to the zeroth-harmonic unpolarized coefficient $A_0^{\mathrm{unp.}}$, thereby displaying their angular patterns and magnitudes relative to the unpolarized baseline. As the $n=0$ coefficient, $A_0^{\mathrm{unp.}}$ carries no azimuthal dependence, so the angular structure of the ratio is determined entirely by the numerator. As dictated by \cref{e.Sivers_selection_rule}, the even harmonics shown here ($n=0,\, 2$) reproduce the canonical Sivers left-right asymmetry, whereas the odd harmonic ($n = 1$) is left-right symmetric.
This is a clear manifestation that soft-gluon azimuthal dependence reshapes the angular structure of TMD observables. In the bottom right panel of \cref{f.Sivers_heat_map}, we sum the harmonics through $n = 3$ to obtain the azimuthal pattern without separating individual harmonics. The higher harmonics induce a modest but appreciable deviation from the $n = 0$ left-right asymmetric baseline, which would be missed if one ignored the azimuthal angular dependence in the soft functions. Consequently, while the individual harmonic series offer novel observables to probe the Sivers effect, incorporating the exact azimuthal dependence of soft radiation is mandatory for robust global analyses of relevant experimental data.

\begin{figure}
\centering
\includegraphics[width = 0.98 \columnwidth]{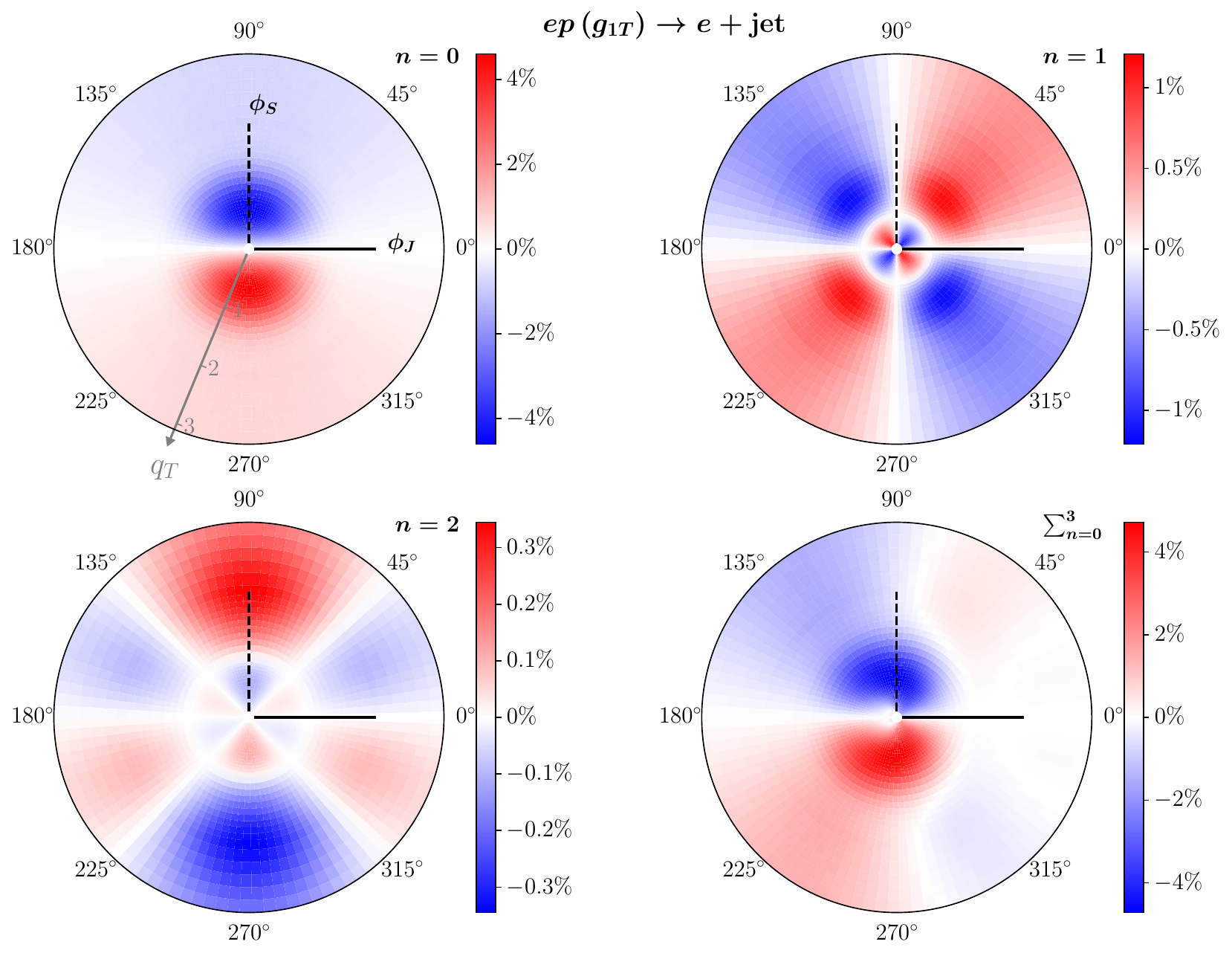}
\caption{Same plot as \cref{f.Sivers_heat_map}, but for the worm-gear asymmetry.}
\label{f.g1T_heat_map}
\end{figure}

In \cref{f.g1T_heat_map}, we map the corresponding worm-gear harmonics $S_T\cos(\phi_q - \phi_S) \cos(n \phi_{qJ}) A_n^{\textrm{worm-gear}} / A_0^{\mathrm{unp.}}$ with $\phi_J = 0$ and $\phi_S = \pi / 2$. As dictated by \cref{e.wormgear_selection_rule}, the canonical up-down asymmetry survives at every harmonic, while the soft radiation generates an additional left-right structure: the even harmonics shown here ($n = 0,\, 2$) are left-right \textit{symmetric}, whereas the odd harmonic ($n = 1$) is left-right \textit{asymmetric}. This odd-harmonic asymmetry has no counterpart in the standard SIDIS worm-gear asymmetry. In the bottom right panel of \cref{f.g1T_heat_map}, we again sum the harmonics through $n = 3$. The deviation from the $n = 0$ up-down asymmetric baseline is more pronounced, underscoring the role of the soft-function angular dependence in extractions of polarized TMD PDFs.

\textit{Conclusions.---}%%
In this work, we study the azimuthal anisotropy of the back-to-back lepton-jet production in lepton-proton collisions.
This process involves three azimuthal angles: those of the transverse spin $\boldsymbol{S}_T$ of the incident proton, the transverse momentum imbalance $\boldsymbol{q}_T$ between the lepton and the jet, and the jet transverse momentum $\boldsymbol{p}_T$. We derive the parity selection rule governing the dependence on these angles within TMD factorization.

Our analysis reveals that the presence of the observed jet breaks the azimuthal isotropy of the soft radiation, introducing a nontrivial angular dependence into the soft function.
This additional structure, which is absent in inclusive processes, significantly complicates the extraction of both the Sivers and worm-gear functions by generating higher-harmonic contributions that can mimic, counteract, or even invert the conventional asymmetry patterns. The even Sivers harmonics retain the canonical left-right asymmetry, whereas the odd harmonics are left-right symmetric. By contrast, the worm-gear function retains its up-down asymmetry at every harmonic, while its even and odd harmonics are left-right symmetric and asymmetric, respectively. The odd worm-gear harmonics thus generate a left-right asymmetry with no counterpart in the standard SIDIS worm-gear asymmetry.

These rich angular patterns are large enough to be measured individually at EIC kinematics, providing new observables of the azimuthal dynamics of soft radiation. A complete treatment of this soft-gluon angular dependence is therefore essential for precision extractions of polarized TMD PDFs.
More broadly, the observables identified in this work offer a new window into the intricate dynamics of soft radiation in QCD, positioning lepton-jet correlations as a powerful tool for exploring the three-dimensional structure of the nucleon in the upcoming era of high-luminosity collider experiments.

We thank Roli Esha, Kyle Lee and Ding Yu Shao for useful discussions. H.X. acknowledges support from the National Natural Science Foundation of China (Grant Nos.~12525508, 12475139). Z.K. is supported by the National Science Foundation under grant No.~PHY-2515057.

%\bibliographystyle{h-physrev5}   
%\bibliography{refs.bib}

\end{document}